\documentclass[doublecol,linenumbers]{epl2} 
\usepackage{enumerate}
\usepackage{graphicx}
\usepackage{amsmath}
\usepackage{amsfonts}
\usepackage{amssymb}
\usepackage{color}
 
\newcommand{\ket}[1]{\left| #1 \right\rangle}
\newcommand{\bra}[1]{\left\langle #1 \right|}

\title{Universal non-adiabatic geometric manipulation of pseudo-spin charge qubits}
\shorttitle{Non-adiabatic, universal and geometric  manipulation of pseudo-spin charge qubits} 

\author{Vahid Azimi Mousolou \inst{1}}

\institute{                    
  \inst{1} Department of Mathematics, Faculty of Science, University of Isfahan, Box 81745-163 Isfahan, Iran
}

\abstract{
Reliable quantum information processing requires high fidelity universal manipulation of quantum systems within the characteristic coherence times. Non-adiabatic holonomic quantum computation offers a promising approach to implement fast, universal, and robust quantum logic gates particularly useful in nano-fabricated solid-state architectures, which typically have short coherence times. Here, we propose an experimentally feasible scheme to realize high-speed universal geometric quantum gates in nano-engineered pseudo-spin charge qubits. We use a system of three coupled quantum dots containing a single electron, where two computational states of a double quantum dot charge qubit interact through an intermediate quantum dot. The additional degree of freedom introduced into the qubit makes it possible to create a geometric model system, which allows robust and efficient single-qubit rotations through careful control of the inter-dot tunneling parameters. We demonstrate that a capacitive coupling between two charge qubits permits a family of non-adiabatic holonomic controlled two-qubit entangling gates, and thus provides a promising procedure to maintain entanglement in charge qubits and a pathway toward fault-tolerant universal quantum computation. We estimate the feasibility of the proposed structure by analyzing the gate fidelities to some extent.}

\begin{document}

\maketitle

\section{Introduction}
In the past few decades, quantum computing has developed from purely abstract ideas into more realistic and practical areas of quantum mechanics. Nevertheless, there are still several considerably big challenges in building a large-scale quantum computer. Among them, short coherence time and fragile nature of qubits require high-speed implementations of quantum gates resilient to certain types of noises. A key approach toward fault-resistant quantum computation is to use geometric (Berry) phase effects \cite{berry84, wilczek84, aharonov87, anandan88} known as holonomic quantum computation (HQC) \cite{Pachos2001, sjoqvist2008, Solinas2012}. The idea of HQC, originally proposed by Zanardi and Rasetti \cite{zanardi99} based on the Wilczek-Zee holonomy \cite{wilczek84}, which is a non-Abelian (non-commuting) extension of  the Berry phase \cite{berry84} associated with adiabatic evolution. Today, HQC appears in different forms:  adiabatic \cite{zanardi99, ekert00} and non-adiabatic \cite{xiang-bin01, sjoqvist12, mousolou14} HQC. The adiabatic condition imposes long gate operation time, which may have a negative impact on running any fast quantum algorithm, whereas a remarkable feature of the non-adiabatic HQC is that there is no major limitation on the gate operation time \cite{erik16}. It was shown in Refs. \cite{sjoqvist12, mousolou14} that speed and robustness can be combined using non-adiabatic non-Abelian quantum holonomies \cite{ anandan88} to perform fast, universal, and robust quantum information processing. Experimentally, non-adiabatic HQC has been  realized in different physical settings, such as NMR \cite{Feng2013}, superconducting artificial atoms   \cite{abdumalikov13}, and NV centers in diamonds \cite{Arroyo-Camejo2014}.
Non-adiabatic HQC has also been combined conceptually with decoherence free subspaces \cite{xu2012}, noiseless subsystems \cite{zhang2014}, and dynamical decoupling \cite{xu2014}. The robustness of non-adiabatic HQC against some general sources of errors has been studied in Ref. \cite{johansson2012}. It is worth mentioning that a different approach has been employed in Ref. \cite{song2016} to speed up HQC.

Scaling of quantum computer architectures to arbitrary numbers of qubits is of central importance and undoubtedly a challenging task. The major challenge is the combination of speed, accuracy and complexity. Namely, implementing a desired large-scale quantum computer requires to perform sufficiently high number of operations within the characteristic coherence time of the qubits with error probabilities below the threshold value needed for efficient fault-tolerant quantum computation in complex quantum systems. The systems, which allow for controllable interactions to create and maintain superposition and entanglement in an arbitrarily big set of well-defined qubits. A natural route towards this end is to look for implementations of non-adiabatic HQC in solid-state devices.
Semiconductor quantum dot based devices, which have great potential to be controlled electrically and scaled within current technologies, are promising candidates. Particular designed systems are based on charge \cite{kouwenhoven_2001, fujisawa06, van-der-Wiel2002} or spin  \cite{loss1998, hanson_2007} degrees of freedom. 
Non-adiabatic control, which depends exclusively on the electric field control, has been demonstrated in GaAs double quantum dot (DQD) charge qubit \cite{petersson_2010, cao2016}. 

In this Letter, we propose an experimentally feasible scheme to realize non-adiabatic HQC with a nano-engineered system of pseudo-spin charge qubits. The geometric qubit structure is designed by a single electron hopping between three quantum dots coupled in a triangular ring geometry. In this scheme, two of the dots, which represent a DQD charge qubit, are coupled to a third auxiliary quantum dot. The additional dot opens up for experimental realization of universal high-speed geometric single-qubit rotations by a particular tuning of inter-dot tunneling parameters. The universality is achieved by developing a family of fully geometric two-qubit entangling gate through an electrostatic interaction between two qubits. The robustness of the proposed structure is illustrated to some extent, showing the feasibility of the structure implementation with current or near future technology. 
The present approach is important both experimentally and theoretically. 
It is non-adiabatic and compatible with short coherence time of quantum dot charge qubits, and thus opens up for experimental realization of holonomic gates with quantum dot charge qubits. It provides a suitable model to explore the interplay between geometric phases and specific features of the quantum dot charge qubit system. This would improve our understanding of the concept of geometric phase/holonomy and its applications in modern physics.

\section{Non-adiabatic quantum holonomy and holonomic gate}
\label{Quantum Holonomy}
Sometimes, only a part of a quantum system is objective because of, for instance, a practical purpose such as information processing.  
Then the states of concern belong to a $m$-dimensional subspace $\mathcal{H}_{m}$ of the corresponding system Hilbert state space $\mathcal{H}$ of dimension $n$. In such a case, we may consider the Schr\"odinger  time evolution
\begin{eqnarray}
\mathcal{C}: [0,\tau]\ni t\mapsto \mathcal{H}_{m}(t),
\end{eqnarray}
started at $\mathcal{H}_{m}(0)=\mathcal{H}_{m}$. Here, each time dependent state $\ket{\psi(t)}\in\mathcal{H}_{m}(t)$ satisfies the time-dependent Schr\"odinger equation 
\begin{eqnarray}
 i\hbar\frac{d}{dt}\ket{\psi(t)}=H^{\text{eff}}(t)\ket{\psi(t)}
\end{eqnarray}
for an initial state $\ket{\psi(0)}\in\mathcal{H}_{m}$, where $H^{\text{eff}}(t)$ is the effective time-dependent Hamiltonian of the system.
The path $\mathcal{C}$ resides in the Grassmannian manifold $G(n, m)$, being the set of all $m$-dimentional subspaces in the $n$-dimensional Hilbert space $\mathcal{H}$. There is a natural bijection between $G(n, m)$ and the set of all projection operators on $\mathcal{H}$ of rank $m$. Suppose $\mathcal{H}_{m}(t)$ undergoes a cyclic evolution, i.e., there is a time $\tau$ such that $\mathcal{H}_{m}(0)=\mathcal{H}_{m}(\tau)$. Note that the cyclic condition on $\mathcal{C}$ does not imply that each state $\ket{\psi(t)}$ evolves about the loop $\mathcal{C}$ in a cyclic manner, in fact we have $\ket{\psi(0)}\neq\ket{\psi(\tau)}$ in general. In order to see how the initial state $\ket{\psi(0)}$ evolves into the final state $\ket{\psi(\tau)}$ along $\mathcal{C}$, it would be sufficient to see how a given initial orthonormal basis (frame) of $\mathcal{H}_{m}$ evolves to a final orthonormal basis of $\mathcal{H}_{m}$ along the closed path $\mathcal{C}$. This follows from the fact that $\mathcal{H}_{m}(t)$ is a linear space. 

Let $B(t)=\left\{\ket{\psi_l(t)},l=1,...,m\right\}$ be a time-dependent family of orthonormal bases given by the time-dependent Schr\"odinger equation $i\hbar\frac{d}{dt}\ket{\psi_l(t)}=H^{\text{eff}}(t)\ket{\psi_l(t)}$ along the path $\mathcal{C}$ for an initial choice of orthonormal basis. Since $\mathcal{C}$ is closed, both $B(0)$ and $B(\tau)$ are orthonormal bases of the same linear space $\mathcal{H}_{m}(\tau)=\mathcal{H}_{m}(0)=\mathcal{H}_{m}$, and therefore they differ only by a unitary transformation, i.e.,
\begin{eqnarray}
\ket{\psi_l(\tau)}=U(\tau)\ket{\psi_l(0)},\ \ \ \ \ \ l=1,...,m,
\end{eqnarray}
where $U(\tau)$ is a $m\times m$ unitary operator. Actually, $U(\tau)$ is the final time evolution operator $\mathcal{U}(\tau, 0)=\exp[-\frac{i}{\hbar}\int^{\tau}_{0}H^{\text{eff}}(t)dt]$ projected onto the subspace $\mathcal{H}_{m}$.
In order to evaluate the unitary operator $U(\tau)$, one may assume another once differentiable family of orthonormal bases $\tilde{B}(t)=\left\{\ket{\zeta_l(t)},l=1,...,m\right\}$ along the $\mathcal{C}$ with $\ket{\zeta_l(\tau)}=\ket{\zeta_l(0)}=\ket{\psi_l(0)}$ for each $l$ and obtain \cite{anandan88} 
\begin{eqnarray}
U(\tau)=\textbf{T}\exp(i\int_0^\tau(\textbf{A}(t)-\textbf{D}(t))dt).
\label{U(tau)}
\end{eqnarray}
Here, 
\begin{eqnarray}
\textbf{A}_{kl}(t)&=&i\bra{\zeta_k(t)}d/dt\ket{\zeta_l(t)}\nonumber\\
\textbf{D}_{kl}(t)&=&\frac{1}{\hbar}\bra{\zeta_k(t)}H^{\text{eff}}(t)\ket{\zeta_l(t)}
\label{phase-potentials}
\end{eqnarray}
are hermitian $m\times m$ matrices, and $\textbf{T}$ denotes the time ordering operator. Under the gauge transformation $\ket{\zeta_k(t)}\rightarrow\sum_{l=1}^kV_{lk}(t)\ket{\zeta_l(t)}$, where $V(t)$ is a once differentiable family of $m\times m$ unitary operators, the matrix $\textbf{A}$ transforms as a proper vector potential according to 
\begin{eqnarray}
\textbf{A}\rightarrow V^{\dagger}\textbf{A}V+iV^{\dagger}\dot{V}
\label{gauge-transformation}
\end{eqnarray}
while $\textbf{D}$ transforms as $\textbf{D}\rightarrow V^{\dagger}\textbf{D}V$. This indicates that the unitary $U(\tau)$ is composed of two parts: the geometric part, which is given by the gauge potential $\textbf{A}$, and the dynamical part, which depends exclusively on the system effective Hamiltonian and is given by the phase factor $\textbf{D}$ \cite{anandan88, bohm2003, dariusz2004}.

Eq. (\ref{phase-potentials}) shows that if the effective Hamiltonian remains trivial along the path $\mathcal{C}$ or equivalently 
\begin{eqnarray}
P_{m}H^{\text{eff}}(t)P_{m}\equiv 0\ \ \ \ \forall t\in [0, \tau], 
\label{zero-dynamical-phase}
\end{eqnarray}
where $P_{m}$ is the projection operator on $\mathcal{H}_{m}$, $\textbf{D}(t)\equiv 0$ and in particular the unitary operator $U(\tau)$ is a purely geometric object. This follows from $[H^{\text{eff}}(t), \mathcal{U}(t, 0)]=0$. In this case, the $U(\tau)$ can be written as the following closed path integral 
\begin{eqnarray}
U(\tau)=U(\mathcal{C})=\textbf{P}e^{i\oint_\mathcal{C}\mathcal{A}},
\end{eqnarray}
where $\mathcal{A}_{kl}=i\bra{\zeta_k}d\ket{\zeta_l}$  defines a matrix-valued connection one-form on the principal bundle $(S(n,m), G(n,m), \pi, U(m))$. The Stiefel manifold $S(n,m)$ is the set of all orthonormal $m$-frames in $\mathcal{H}$, $\pi$ is the natural projection that maps each $m$-frame to the corresponding $m$-dimensional subspace spanned by that frame, and $U(m)$ is the group of unitary operators of rank $m$ \cite{bohm2003, dariusz2004}. Moreover, the path $C:[0, \tau]\ni t\mapsto B(t)$ in $S(n, m)$ is the horizontal lift of $\mathcal{C}$  with respect to the connection form $\mathcal{A}$, and the unitary operator $U(\mathcal{C})$ is the holonomy of this lift \cite{bohm2003, dariusz2004}. The quantum holonomy $U(\mathcal{C})$ is the non-adiabatic non-Abelian generalization of the quantum geometric phase \cite{anandan88}, which is known as Aharonov-Anandan phase for the special Abelian case $m=1$ \cite{aharonov87}. In the adiabatic limit, this corresponds to the Wilczek-Zee holonomy \cite{wilczek84} for $m\ge 2$ and the Berry phase \cite{berry84} for the special case $m=1$. It is important to stress that although the adiabatic geometric phases \cite{berry84, wilczek84} occur in slow evolutions of energy eigensubspaces, in the non-adiabatic holonomy $U(\mathcal{C})$ the subspace $\mathcal{H}_{m}$ needs to be neither an energy eigensubspace nor evolved slowly.  Further conceptual aspects of geometric phase/holonomy specially in the context of quantum computation can be found in Ref. \cite{erik16}. 

Note that geometric phases and quantum holonomies are global properties of quantum evolutions and depend only on the geometric structure of the state space \cite{bohm2003, dariusz2004}. Thus, they are inherently resilient to local perturbations, external parameter noises, and some other class of errors associated with specific details of how the evolutions are carried out \cite{Pachos2001, sjoqvist2008, Solinas2012}. Namely, quantum holonomies have some built-in fault-tolerant features and stability, which can be employed to achieve robust quantum gates for quantum computation. Holonomic gates based on non-adiabatic non-Abelian geometric phase offer some additional advantages: they are universal, they are exact in a sense that there is no adiabatic approximation, and there is no need for slow manipulation that implies they would be much faster than their adiabatic counterpart. The latter can be particularly useful in solid-state devices, where qubits typically have short coherence times, including pseudo-spin charge qubits.
 
In the following, we consider a model system that can be used to realize a universal set of non-adiabatic holonomic gates with pseudo-spin charge qubits.  

\section{Model system}
\begin{figure}[h]
\centering
\includegraphics[width=65mm,height=50mm]{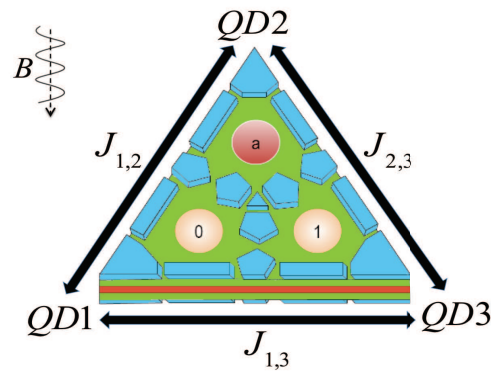}
\caption{(Color online) Basic device structure for the physical realization of non-adiabatic HQC. A set of metallic gates are used to construct three quantum dots arranged in a triangular ring geometry out of a semiconductor heterostructure. It is assumed that this three-quantum-dot device contains only a single electron. Single-qubit holonomic gates are realized when the system is exposed to an external magnetic field $\textbf{B}$ and the hopping between first and third sites is completely suppressed, i.e. $J_{1,3}=0$. During a computation the barriers between adjacent quantum dots are controlled by the same time-dependent scaling function, i.e. $J_{k,k+1} = \Omega (t) \mathcal{J}_{k,k+1}$, via electric gates. } 
\label{fig:ring-geometry}
\end{figure}

The system that we have in mind is an artificially constructed lateral quantum dot \cite{van-der-Wiel2002, hanson_2007} by depleting the well-known two-dimensional electron gas supported, e.g., by a GaAs/AlGaAs or a  Si/SiGe heterostructure via a set of metallic gates that can be filled with electrons or holes. We assume an appropriate set of metal gate electrodes on the heterostructure surface to form three quantum dots arranged in a triangular ring geometry as shown in Fig. \ref{fig:ring-geometry}.  State of the art electron beam lithography is a fundamental technique that has been frequently employed for the fabrication of this type of quantum confinement in a semiconductor heterostructure \cite{muhle2008, petersson_2010, ward_2016}. In this setup, quantum dots are located at a proper distance from each other such that electrons can tunnel between the neighboring dots in a controllable manner. The tunnel coupling strength between adjacent quantum dots can be controlled individually by tuning the inter-dot potential barrier through metal gates. Single-particle energy levels in each quantum dot are assumed to be controlled and individually manipulated with prepared gate electrodes on the back of the heterostructure. Non-adiabatic HQC can be realized when this three-quantum-dot device contains only a single electron. This single-electron tunneling device forms the basic qubit building block in what follows. As shown in Fig. \ref{fig:ring-geometry}, our qubit in a way is a DQD charge qubit  represented by dots at the first and third sites, which are coupled through an auxiliary dot at the second site. In the following, the auxiliary quantum dot provides an extra degree of freedom, which allows geometric manipulation of the DQD charge qubit state with great speed and robustness. A different approach has been considerd in \cite{Shi_2012} to speed up operations on DQD charge qubit by introducing more degrees of freedom into the qubit architecture.

If we restrict ourselves to the case of single-particle orbital in each quantum dot, the  three-quantum-dot system can be described by a three-site tight-binding model. The system Hamiltonian in the second quantization language for a spinless electron reads

\begin{equation}
H_{\text{q}} = \sum_{k=1}^3 \epsilon_{k}c_{k}^{\dagger} c_{k}^{\phantom\dagger} + 
\sum_{k=1}^3 \left[t_{k,k+1} c_{k} ^{\dagger}c_{k+1}^{\phantom\dagger}+h.c.\right],
\label{eq:hamiltonian1}
\end{equation}
where  $t_{k,k+1}$ are nearest-neighbor hopping matrix elements, which can be controlled by gate voltages. 
The operators $c_{k}^{\dagger}$ ($c_{k}^{\phantom\dagger}$) create (annihilate) electrons in the corresponding Wannier states localized at dot-site $k$, with on-site energy $ \epsilon_{k}$. We assume periodic boundary condition such that the first and fourth sites are identified. 

To exploit the full computational potential of our architecture, we assume that the system is exposed to an external magnetic field $\textbf{B}$.   An appropriate way of dealing with magnetic fields in a tight-binding model is given by the Peierls substitution method \cite{graf95}. In this way, the influence of the magnetic field is described by gauge-invariant modification of the tight-binding hopping parameters via the Peierls phase factors 

\begin{eqnarray}
t_{k,k+1} \rightarrow  J_{k,k+1}=t_{k,k+1}e^{-i\frac{e}{\hbar} {\alpha_{k,k+1}}} ,
\label{eq:hopping}
\end{eqnarray}
where $\alpha_{k,k+1}=\int_{k\rightarrow k+1} \mathbf{A} \cdot d \mathbf{l}$ 
is the integral of the vector potential along the hopping path from site $k$ to site $k+1$. The Hamiltonian $H_{\text{q}}$ in Eq. (\ref{eq:hamiltonian1}) with complex valued hopping matrix elements $ J_{k,k+1}$ in Eq. (\ref{eq:hopping}) describes the dynamic of our qubit system. 

\section{Single-qubit gates}
We pursue by fixing the on-site energies $\epsilon_{k}=0 $ via adjusting back-gate voltages on the device. We assume that hopping between first and third sites is completely suppressed, i.e. $J_{1,3}=0$, by increasing the tunnel barrier between the corresponding dots. We further assume that hopping parameters are turned on and off with the same time-dependent scaling function $\Omega (t)$ by sweeping the top gates. More explicitly speaking, we assume $J_{k,k+1} = \Omega (t) \mathcal{J}_{k,k+1}$ with time-independent complex-valued coupling parameters $\mathcal{J}_{k,k+1}$ for each $k$. It is important to stress that this can be implemented by efficiently manipulating the inter-dot tunnel couplings $t_{k,k+1}$ with time-dependent electric gate voltages. 
In the Fock space spanned by the three one-electron states $\{\ket{100},\ket{010},\ket{001}\}$, the effective time-dependent Hamiltonian is given by ($ \hbar= 1$ from now on)
\begin{eqnarray}
H_{\text{q}}^{\text{eff}}(t) = \Omega(t)[ \mathcal{J}_{1,2}^{*}\ket{010}\bra{100}+ \mathcal{J}_{2,3}\ket{010}\bra{001}+h.c.].\nonumber\\
\label{eq:hamiltonian}
\end{eqnarray}
We notice that, the Hamiltonian in Eq.~(\ref{eq:hamiltonian}) has the same form as the Hamiltonian for the three-level $\Lambda$ configuration in Ref. \cite{sjoqvist12} if we encode the one-electron states as 

\begin{eqnarray}
\ket{0}:=\ket{100},\ \ \ \ \ \ket{a}:=\ket{010}\ \ \ \  \text{and}\ \ \ \  \ket{1}:=\ket{001}\ \ \ \ 
\label{eq:basisencoding}
\end{eqnarray}
Hence, the DQD charge qubit is defined by whether the electron is localized to the first dot or the third (see Fig. \ref{fig:ring-geometry}).
 
Now, a universal set of holonomic one-qubit gates can be realized if we consider evolutions of the qubit subspace, $\mathcal{H}_{q}=\text{span}\{\ket{0}, \ket{1}\}$, by the effective Hamiltonian in Eq.~(\ref{eq:hamiltonian}) about closed paths in the space of two-dimensional subspaces of  the three-dimensional Hilbert space $\mathcal{H}=\text{span}\{\ket{0}, \ket{a}, \ket{1}\}$, i.e., the Grassmann manifold $G(3,2)$. One way to achieve such evolutions, is to turn on and off the Hamiltonian in Eq. (\ref{eq:hamiltonian}) during the time interval $[0, \tau]$, controlled by the function $\Omega(t)$ under the condition that $\sqrt{|\mathcal{J}_{1,2}|^{2}+|\mathcal{J}_{2,3}|^{2}}\int_{0}^{\tau}\Omega(t)dt=\pi$. 
Explicitly, if we choose the coupling constants $\mathcal{J}_{1,2}^{*}=\sin(\theta/2)e^{i\phi}$ and $\mathcal{J}_{2,3}=-\cos(\theta/2)$, the qubit space $\mathcal{H}_{q}$
undergoes a cyclic evolution  in $G(3,2)$, where $\int_{0}^{\tau}\Omega(t)dt=\pi$. In this case, the final time evolution operator $\mathcal{U}(\tau, 0)$ projected onto the qubit subspace, $\mathcal{H}_{q}$, is given by 

\begin{eqnarray}
U(\mathcal{C})={\mathbf n}.{\mathbf\sigma}
\label{eq:gate}
\end{eqnarray}
in the ordered computational qubit basis $\{\ket{0},\ket{1}\}$. Here ${\mathbf n}=(\sin\theta\cos\phi, \sin\theta\sin\phi, \cos\theta)$, $\mathbf{\sigma}=(\sigma_{x},\sigma_{y},\sigma_{z})$ is the Pauli matrix vector, and $\mathcal{C}$ is the closed path in $G(3,2)$ specified by parameter angles $\theta$ and $\phi$, about which the qubit subspace evolves. The  $\Lambda$  form of the system effective Hamiltonian, $H_{\text{q}}^{\text{eff}}(t)$, guarantees that  $\bra{l}H_{\text{q}}^{\text{eff}}(t)\ket{k}=0$ for each $t\in [0, \tau]$, $k,l=0,1$. In other words, $P_{q}H_{\text{q}}^{\text{eff}}(t)P_{q}=0$ for each $t\in [0, \tau]$, where $P_{q}$ is the projection operator on $\mathcal{H}_{q}$. This indicates, according to Eq. (\ref{zero-dynamical-phase}), that the dynamical phase accompanying the cyclic evolution of  $\mathcal{H}_{q}$ about $\mathcal{C}$ vanishs. Consequently, the unitary operator $U(\mathcal{C})$ is the non-adiabatic non-Abelian quantum holonomy associated with $\mathcal{C}$, and thus is solely geometric and fully determined by the closed path $\mathcal{C}$ in the Grassmann manifold $G(3,2)$.
 
Note that the magnetic field is fixed constant during the implementation of the holonomic gate $U(\mathcal{C})$.
For a given magnetic field, we must choose the complex-valued hopping parameters given in Eq. (\ref{eq:hopping}) so that 
their phases $\alpha_{k,k+1}$ add up to the  total magnetic flux $\Phi = \oint \mathbf{A} 
\cdot d \mathbf{l}= {\cal A} B$, where $\cal A$ is the area enclosed by the three-dot ring. 
A change of these phases corresponds to a gauge transformation $\mathbf{A} \rightarrow 
\mathbf{A} + \nabla\Pi$ under which the holonomy is transformed into $U(\mathcal{C})\rightarrow\Gamma^{\dagger} U(\mathcal{C}) 
\Gamma$, where $\Gamma = {\textrm{diag}} \{e^{-i\frac{e}{\hbar} 
\Pi(1)}, e^{-i\frac{e}{\hbar}\Pi(3)}\}$ with $\Pi(k)$ indicating the 
value of $\Pi$ at site $k$. Thus, the holonomies associated with this 
field-dependent tight-binding Hamiltonian are gauge covariant quantities. 
Further conceptual facts about the geometric nature of $U(\mathcal{C})$ are discussed in \cite{erik16}. 
 
  The $U(\mathcal{C})$ is universal in the sense that for a set of parameter angles $\theta$ and $\phi$ it generates the whole special unitary group $SU(2)$, explicitly,
  
\begin{eqnarray}
U(\mathcal{C})U(\tilde{\mathcal{C}})=({\mathbf n}.{\mathbf\sigma}) ({\mathbf m}.{\mathbf\sigma})=\mathbf n\cdot\mathbf m-i\mathbf\sigma(\mathbf n\times \mathbf m),
\end{eqnarray}
has the general form of an $SU(2)$ transformation. In other words, the $U(\mathcal{C})$ is a universal geometric one-qubit gate, which allows for universal quantum information processing if it is accompanied by a two-qubit entangling gate operator.

\begin{figure}[t]
\centering
\includegraphics[width=50mm,height=25mm]{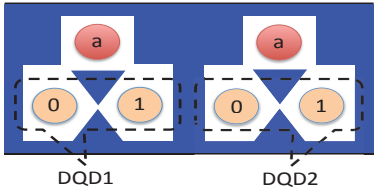}
\caption{(Color online) Two-qubit model configuration.}
\label{fig:2qubitconfig}
\end{figure}

\section{Two-qubit gate}  
Fig. \ref{fig:2qubitconfig} depicts our structure for a two-qubit system. In this structure, we arrange two of the three-quantum-dot devices, which are described by the Hamiltonian in Eq. (\ref{eq:hamiltonian1}), in the vicinity of each other so that the quadruple quantum dots specifying two charge qubits are all placed along the same line forming a linear geometry (see Fig. \ref{fig:2qubitconfig}).  We assume that charge transition between separate devices are forbidden. The system Hamiltonian reads
\begin{eqnarray}
H_{\text{q-q}} = H_{\text{q}}^{(1)}+H_{\text{q}}^{(2)}+H_{\text{CI}},
\label{eq:hamiltonian2}
\end{eqnarray}
where $i=1, 2,$ labels the two devices in the structure. The first two terms on the right hand side represent single-device system contributions, and the third term originates from the electrostatic Coulomb interaction between excess electrons in the system.

To achieve an effective two-qubit model system, we turn off the local external magnetic fields and concentrate only on charge transition between a pair of quantum dots representing a  charge qubit in each individual device by assuming \cite{note}
\begin{eqnarray}
t^{(i)}_{1,2}=t^{(i)}_{2,3}=0, \ \ \ \ \ \ \ i=1,2.
\end{eqnarray}
In this way, the auxiliary quantum dots are decoupled from the system and the system is reduced to a pair of DQD charge qubits coupled electrostatically in the linear geometry. By appropriate tuning of gate voltages, the system is described effectively by two-qubit Hamiltonian
\begin{eqnarray}
H_{\text{q-q}}^{\text{eff}}=t_{1,3}^{(1)}\sigma_{x}\otimes I+t_{1,3}^{(2)}I\otimes\sigma_{x}+\alpha\sigma_{y}\otimes\sigma_{y}\label{eq:2qubithamiltonian}
\label{eq:2qubiteffeH}
\end{eqnarray}
in the invariant computational subspace spanned by the ordered basis $\{\ket{00}, \ket{01}, \ket{10}, \ket{11}\}$ (see Eq. (\ref{eq:basisencoding})).  Here $I$ is the $2\times 2$ identity matrix and $\alpha$ represents the coupling strength between the two qubits. The qubits interaction, expressed by the third term in Eq. (\ref{eq:2qubiteffeH}), follows from the fact that electron tunneling in the first DQD (control qubit) influences electron tunneling in the second DQD (target qubit) through the Coulomb force between the two excess electrons in the system \cite{note1}.   We neglect other types of qubits interaction terms \cite{fujisawa06, ward_2016, petersson09} on the basis that they have more effect on $\epsilon^{(i)}_{k}$ and $t_{k, j}^{(i)}$ , which can be effectively compensated by tuning gate voltages. 

Now if we further put $t_{1,3}^{(2)}=0$ by tuning the corresponding gate voltage, the Hamiltonian in Eq. (\ref{eq:2qubithamiltonian}) splits the computational two-qubit space into two orthogonal subspaces $\mathcal{H}_{+}=\text{span}\{\ket{00}, \ket{01}\}$ and $\mathcal{H}_{-}=\text{span}\{\ket{10}, \ket{11}\}$ associated with states of the control qubit. This yields the time evolution operator 
\begin{eqnarray}
\mathcal{U}(t,0)  = \cos(\omega t)I-\frac{i}{\omega}\sin(\omega t)H
\label{eq:u2(t)}
\end{eqnarray}
if the hopping parameter $t_{1,3}^{(1)}$ is turned on and off by a time-dependent square pulse envelope $\Phi(t)$ with pulse amplitude 
$\Phi$. Here we have   
\begin{eqnarray}
t_{1,3}^{(1)}&=&\Phi(t)\delta,\ \ \ \ \ \ \ \ \ \ \ \omega=\sqrt{(\Phi\delta)^{2}+\alpha^{2}}\nonumber\\
H&=&
\left(
\begin{array}{cc}
 0 &   T  \\
  T^{\dagger}   &  0  
\end{array}
\right),\ \ \ \ T= (\Phi\delta) I-i\alpha\sigma_{y}\ \ \ 
\end{eqnarray}
with $\delta$ being time-independent coupling parameter. 

A two-qubit entangling gate is realized by first applying a square pulse over the time interval $[0, \tau]$ with pulse amplitude $\Phi$ and time dependent hopping $t_{1,3}^{(1)}=\Phi(t)\delta$, followed by a square pulse over the time interval $[\tau^{\prime}, \tau^{\prime\prime}]$ with pulse amplitude $\tilde{\Phi}$ and time dependent hopping $\tilde{t}_{1,3}^{(1)}=\tilde{\Phi}(t)\delta$. We assume that $\tau<\tau^{\prime}<\tau^{\prime\prime}$ and that the Hamiltonian is completely turned off during $[\tau,\tau^{\prime}]$ or the time gap $\tau^{\prime}-\tau$ is negligibly small. Note that the time gap $\tau^{\prime}-\tau$ is only restricted  
by decoherence and errors and can be arbitrary long in the ideal noise-free case. By choosing parameters such that $\sin(\omega\tau)=(-1)^{n}$ and $\sin(\tilde{\omega}(\tau^{\prime\prime}-\tau^{\prime}))=(-1)^{\tilde{n}}$, where $n, \tilde{n}=0,1$, we obtain the final time evolution operator 
\begin{eqnarray}
\mathcal{U}(\tau^{\prime\prime},0)=\mathcal{U}(\tau^{\prime\prime},\tau^{\prime};\tilde{t}_{1,3}^{(1)})\mathcal{U}(\tau,0;t_{1,3}^{(1)})=
 U(\mathcal{C}_{+})\oplus U(\mathcal{C}_{-}),\nonumber\\
\label{eq:u2(t)os}
\end{eqnarray}
where $ U(\mathcal{C}_{\pm})=(-1)^{n+\tilde{n}+1}\mathcal{R}(\pm\varphi)$ with $\mathcal{R}(\varphi)$ being the $2\times 2$ rotation matrix of angle $\varphi$ such that $\sin(\varphi)=\frac{\alpha\delta(\Phi-\tilde{\Phi})}{\tilde{\omega}\omega}$.

\begin{figure}[h]
\centering
\includegraphics[width=70mm,height=35mm]{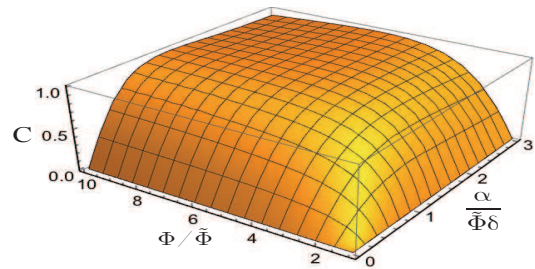}
\caption{ (Color online) The concurrence $C\equiv C[U(\mathcal{C}_{+})\oplus U(\mathcal{C}_{-}]$ as a function of two dimensionless parameters $\Phi/\tilde{\Phi}$ and $\alpha/\tilde{\Phi}\delta$.}
\label{fig:concurrence}
\end{figure}

For any two different pulses, $\Phi-\tilde{\Phi}\neq 0$, the unitary operator in Eq. (\ref{eq:u2(t)os}) is a two-qubit entangling gate with operator entanglement quantified by concurrence $C[\mathcal{U}(\tau^{\prime\prime},0)]=|\sin2\varphi|$ \cite{wang02}. Fig. \ref{fig:concurrence} shows that any entangling power can be achieved by  appropriate choice of pulses. The  entangling nature of $\mathcal{U}(\tau^{\prime\prime},0)$ confirms the universality of our scheme.

A remarkable feature of the entangling gate $\mathcal{U}(\tau^{\prime\prime},0)$  in Eq. (\ref{eq:u2(t)os}) is that it is a purely geometric gate. To see this, one may note that the first pulse evolves the system initialized at the subspace $\mathcal{H}_{\pm}$ along a geodesic, $\gamma_{\pm}$, in the Grassmann manifold $G(4,2)$ into the orthogonal counterpart subspace $\mathcal{H}_{\mp}$, and then the second pulse takes the system at $\mathcal{H}_{\mp}$ back into the initial subspace along a different geodesic, $\tilde{\gamma}_{\pm}$, in $G(4,2)$ \cite{mousolou14b}. In particular, the two subsequent pulses evolve both subspaces $\mathcal{H}_{+}$ and $\mathcal{H}_{-}$, respectively, about the closed orange slice shaped paths $\mathcal{C}_{+}=\gamma_{+}*\tilde{\gamma}_{+}$ and $\mathcal{C}_{-}=\gamma_{-}*\tilde{\gamma}_{-}$ in $G(4,2)$.
Moreover, the block off-diagonal form of the Hamiltonian $H_{\text{q-q}}^{\text{eff}}(t; t_{1,3}^{(2)}=0)$ assures that
\begin{eqnarray}
 P_{+}H_{\text{q-q}}^{\text{eff}}(t; t_{1,3}^{(2)}=0)P_{+}&=&0,\nonumber\\
 P_{-}H_{\text{q-q}}^{\text{eff}}(t; t_{1,3}^{(2)}=0)P_{-}&=&0
 \end{eqnarray}
 at any time $t$, where $P_{\pm}$ are the projection operators on the corresponding subspaces $\mathcal{H}_{\pm}$. Thus, from the discussion including Eq. (\ref{zero-dynamical-phase}) we have that the dynamical phases all vanish along the four paths $\gamma_{\pm}$ and $\tilde{\gamma}_{\pm}$, and consequently along both cyclic evolutions $\mathcal{C}_{\pm}$. This, together with the fact that $U(\mathcal{C}_{\pm})$ are actually the time evolutions of the subspaces $\mathcal{H}_{\pm}$ about the closed paths $\mathcal{C}_{\pm}$ in $G(4,2)$, imply that the unitary operators $U(\mathcal{C}_{\pm})$ are nothing but the non-adiabatic non-Abelian quantum holonomies associated with the loops $\mathcal{C}_{\pm}$. Since $\mathcal{U}(\tau^{\prime\prime},0)$ is the direct sum of the two non-adiabatic non-Abelian quantum holonomies $U(\mathcal{C}_{\pm})$, it is thus a purely geometric gate. For more about the geometric nature of this type of gates, one may see Ref. \cite{erik16}. 
     
\section{Fidelity}

\begin{figure}[h]
\centering
\includegraphics[width=80mm,height=36mm]{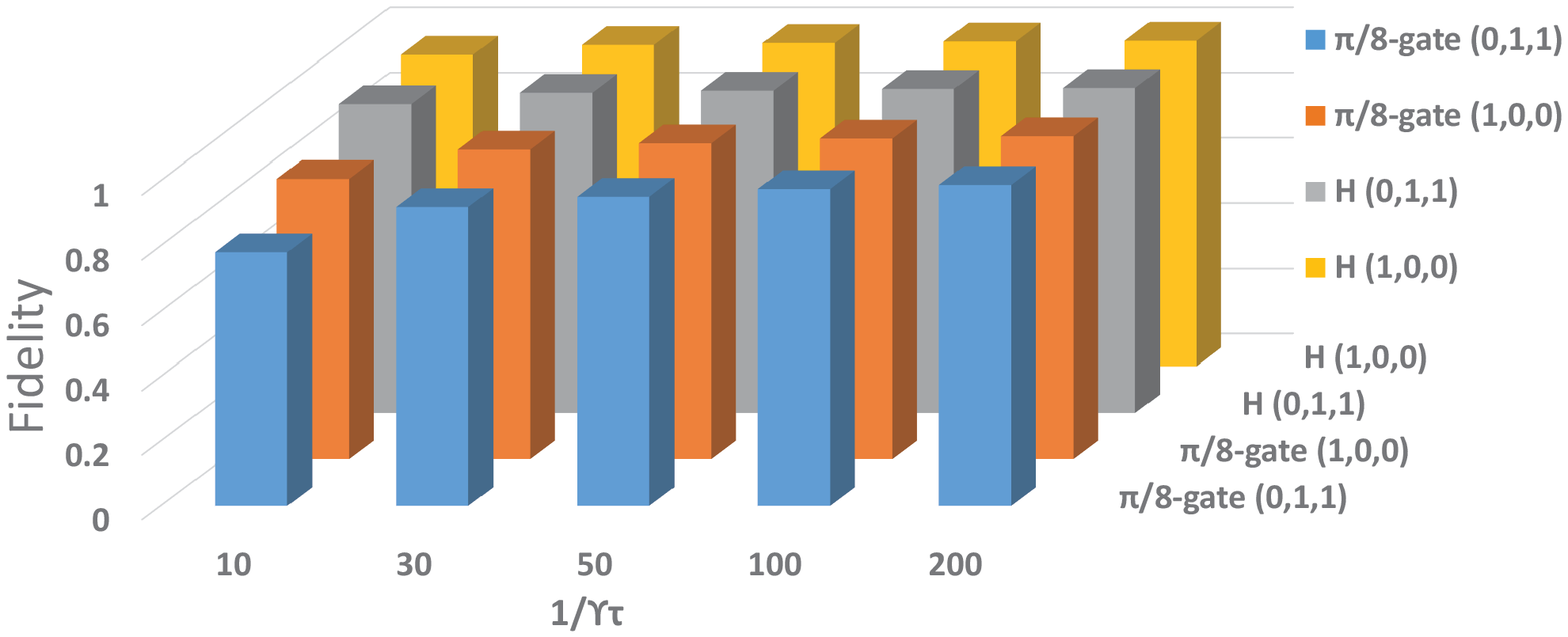}
\includegraphics[width=70mm,height=37mm]{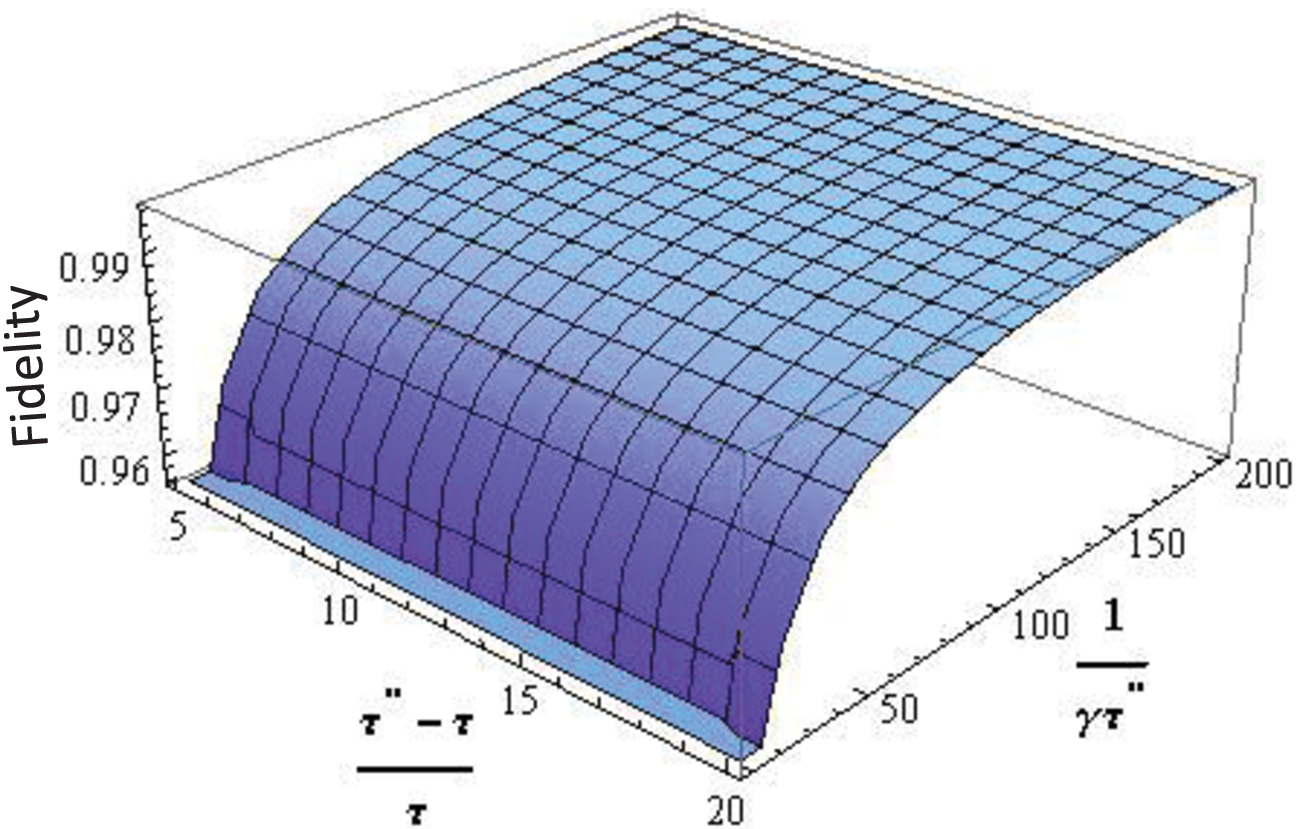}
\caption{(Color online) Influence of decay mechanisms, which cause fluctuation of the energy parameters $\epsilon^{(i)}_{k}$, on a universal set of non-adiabatic holonomic gates. We assume the affected dots in the sample to have the same decay rate $\gamma$. (Upper panel) Fidelity of the holonomic Hadamard and $\pi/8$ gates as a function of  $1/\gamma\tau$, i.e., the ratio between the decay time $1/\gamma$ and the gate oprtational time $\tau$. Triplets of zeros and ones associated with each gate indicate the dots relatively involved in the decay process. (Lower panel) Fidelity of the holonomic two-qubit entangling gate $\mathcal{R}_{c}(\pi/4)=\mathcal{R}(\pi/4)\oplus\mathcal{R}(-\pi/4)$ as a function of $\tau^{\prime\prime}-\tau/\tau$, i.e., the ratio between second and first pulse time intervals, and $1/\gamma\tau^{\prime\prime}$, where $\tau^{\prime\prime}$ is the gate oprtational time. In the two-stage gates $\pi/8$ and $\mathcal{R}_{c}(\pi/4)$, switching times between subsequent pulses are assumed to be negligible.}
\label{fig:fidelity}
\end{figure}

Actual dynamics involves undesirable decoherence processes due to the effect of the electromagnetic environment. We have examined the robustness of our scheme by evaluating the fidelity of a universal set of gates against some important source of decohernce in coupled quantum dot charge qubit systems, such as decay and dephasing \cite{fujisawa06}. If we consider the ratio between the decoherence time and the gate operational time of $>100$, in principle, we can achieve a high fidelity of $>98\%$. Thus, with typical coherence time of the order of $10$ ns \cite{petersson_2010}, high fidelity gates can be implemented within the time scale of $100$ ps. Fig.~\ref{fig:fidelity} shows the fidelity of a universal set of non-adiabatic holonomic gates influenced by a decay process due to, e.g., background charge noise and electrical noise in the gate voltages.

\section{Conclusions}
In conclusion, we have established an experimentally feasible scheme to implement high-speed universal geometric quantum gates on nano-engineered pseudo-spin charge qubits, which require fast coherent manipulation due to their relatively short coherence times. In the case of single-qubit gate, we introduce an extra degree of freedom by attaching an auxiliary quantum dot to the DQD charge qubit in a way that the three dots in the sample form a triangular ring geometry. The additional degree of freedom allows us to construct a geometric $\Lambda$ model system and thus implement any non-adiabatic holonomic single-qubit gate by appropriate electrical control of the inter-dot coupling parameters.  We show that the proposed scheme permits universal quantum information processing by demonstrating a family of  non-adiabatic holonomic two-qubit  entangling gates through an electrostatic interaction between two charge qubits. We further show that this interaction allows any entangling power to be implemented between qubits.
In order to estimate the feasibility, we have examined the robustness of the structure to some extent. We illustrate that if the ratio between the decoherence time and the gate operational time is $>100$, in principle, high fidelity gates can be achieved. Non-adiabatic and geometric nature of the proposed holonomic gates, universality, robustness and feasibility of the structure are feature that distinguish our scheme from previous proposals with charge and spin qubits in quantum dots.

Last but not the least, the short coherence time of quantum dot charge qubits may not allow the realization of adiabatic holonomic quantum computation, but our scheme opens up for experimental realization of non-adiabatic holonomic gates on quantum dot charge qubits. Since each system has its own properties and requirements, it is important that various types of holonomic gates be thoroughly investigated and realized in different physical systems and compared to each other. This would further improve our theoretical and practical understanding of the concept of geometric phase/holonomy in modern physics. It would also help shed light on the role of different types of geometric phases in achieving the most robust, at the same time, the most experimentally accessible way of manipulating quantum systems for information processing. The present paper is a contribution in this direction.

\acknowledgments
The author acknowledges support from Department of Mathematics at University of Isfahan (Iran).


\begin{thebibliography}{0}
\bibitem{note} At removal of the couplings, we assume that the excess electrons are not in the auxiliary dots.
\bibitem{note1} The idea behind the coupling term, $\alpha\sigma_{y}\otimes\sigma_{y}$, is like those discussed in Ref. \cite{fujisawa06} between two charge qubits arranged parallel or perpendicular to each other. In our case, in which the two qubits are arranged in the linear geometry, when the state of the control qubit changes from $\ket{0}$ to $\ket{1}$, the distance between two excess electrons is reduced. In other words, the Coulomb force between the two excess electrons is increased. Therefore, if the target qubit were in the state $\ket{0}$, it would feel more force to change its state to $\ket{1}$, but if the target qubit were in the state $\ket{1}$, it would feel more force not to change its state to $\ket{0}$. So, while the former case induces a positive interaction between $\ket{00}$ and $\ket{11}$, the latter case induces a negative interaction between $\ket{01}$ and $\ket{10}$. There is a similar scenario starting with $\ket{1}$ for the state of the control qubit. This validates the existence of a coupling term like $\alpha\sigma_{y}\otimes\sigma_{y}$ in the linear configuration.
\end{thebibliography}
\end{document}